\documentclass[journal]{IEEEtran}
\usepackage{cite}

\ifCLASSINFOpdf
\usepackage[pdftex]{graphicx}
\usepackage{siunitx}
\usepackage{booktabs}
\usepackage{xcolor}
\def\SB[#1]{\textcolor{blue}{#1}}
\else
\fi
\usepackage{subcaption}
\usepackage[shortlabels]{enumitem}
\usepackage{multirow}
\usepackage{hyperref}
\usepackage{amsmath,amssymb,amsfonts}
\usepackage{diagbox}

\begin{document}

\title{Resampling Filter Design for Multirate Neural Audio Effect Processing}

\author{Alistair~Carson,
    Vesa~Välimäki,
    Alec~Wright,
    and~Stefan~Bilbao
\thanks{A. Carson, A. Wright and S. Bilbao are with the Acoustics and Audio, University of Edinburgh, Edinburgh, UK}
\thanks{V. Välimäki is with the Acoustics Lab, Department of Information and Communications Engineering, Aalto University, Espoo, Finland.}
\thanks{A. Carson is funded by the Scottish Graduate School of Arts and Humanities (SGSAH), e-mail: \href{mailto:acarson2@ed.ac.uk}{alistair.carson@ed.ac.uk}.}
}

%
%

\markboth{}%
{}
%



\maketitle

\begin{abstract}
Neural networks have become ubiquitous in audio effects modelling, especially for guitar amplifiers and distortion pedals. One limitation of such models is that the sample rate of the training data is implicitly encoded in the model weights and therefore not readily adjustable at inference. Recent work explored modifications to recurrent neural network architecture to approximate a sample rate independent system, enabling audio processing at a rate that differs from the original training rate. This method works well for integer oversampling and can reduce aliasing caused by nonlinear activation functions. For small fractional changes in sample rate, fractional delay filters can be used to approximate sample rate independence, but in some cases this method fails entirely. Here, we explore the use of real-time signal resampling at the input and output of the neural network as an alternative solution. We investigate several resampling filter designs and show that a two-stage design consisting of a half-band IIR filter cascaded with a Kaiser window FIR filter can give similar or better results to the previously proposed model adjustment method with many fewer filtering operations per sample and less than one millisecond of latency at typical audio rates. Furthermore, we investigate interpolation and decimation filters for the task of integer oversampling and show that cascaded half-band IIR and FIR designs can be used in conjunction with the model adjustment method to reduce aliasing in a range of distortion effect models.
\end{abstract}


%
\IEEEpeerreviewmaketitle
\section{Introduction}
\IEEEPARstart{V}{irtual} analog modelling refers to the emulation of analog audio effects using digital signal processing, allowing users to replace analog hardware with software \cite{DAFXVAchapter}. Of particular interest is the modelling of guitar amplifiers and distortion effects \cite{Damskaag2018, Parker2018, juvela2023end, Buffa23}. In recent years, neural network based modelling techniques have seen extensive investigation \cite{Wright2019RNN, Wright2020, juvela2023end, vanhatalo2022review}. Several products claiming to use neural technology are now on the market, and the availability of open-source implementations has led to the development of several online databases of crowd-sourced models \cite{GuitarML, NAM, Wright2025}. Recurrent neural networks (RNNs) are a popular choice of model and can produce perceptually convincing, real-time models of tube amplifiers and distortion pedals \cite{Wright2020, wright2020_perceptual, cassidy2023perceptual}.

A limitation of RNNs is that the model sample rate is determined by that of the training data and is not readily adjustable at inference time. This may pose a problem when a user wishes to process audio sampled at a different rate (for example if the model was trained on audio at \SI{44.1}{\kHz} but the input signal at inference is sampled at \SI{48}{\kHz}). Previous work investigated and proposed solutions to this problem by implementing fractional delay filters within the RNN feedback loop \cite{Chowdhury2022, Carson2024, Carson2025}. We refer to these models as sample rate independent RNNs (SRIRNNs). This method can provide good results when the inference sample rate is greater than that of training, and can be exploited to reduce aliasing via oversampling \cite{Carson2024, Carson2025}. When the inference sample rate is lower than the training rate, however, this method requires the prediction of a fractional signal advance which can cause model instability and audible artifacts in the output signal \cite{Carson2025}. 
Sample rate independence has also been investigated for convolutional neural networks (CNNs) for audio source separation \cite{Saito2022}, but the application to audio effect CNNs (e.g. \cite{Damskaag2018, steinmetz2022, Comunita2022}) is outwith the scope of this work.

In this work, we explore the option of operating RNNs at the originally intended sample rate via signal resampling. This involves first resampling the input signal from the input rate to the model rate, RNN processing at the model rate, and then resampling the output signal from the model rate back to the original input sample rate. We investigate several resampling filters for this task, and compare the results against the SRIRNN approach mentioned in the previous paragraph. The suitability of both methods for real-time audio processing is evaluated through measures of latency and number of operations per sample.

Furthermore, we explore the combined use of resampling filters with the SRIRNN method to reduce aliasing via oversampling\cite{Carson2024}. Here, we investigate the best resampling filters for interpolation before and decimation after RNN processing. 

This paper is structured as follows: Sec. \ref{sec:rnns} provides an overview of RNNs for audio processing and the SRIRNN method; Sec. \ref{sec:resampling} outlines the design approach for the resampling filters; Sec. \ref{sec:metrics} describes the evaluation method; Sec. \ref{sec:CD_DAT} investigates the problem of a fractional discrepancy between the model rate and inference rate; Sec. \ref{sec:oversampling} concerns oversampling of RNNs for aliasing reduction and the resampling filters for this task; Sec. \ref{sec:conclusion} provides concluding remarks and recommendations.
Accompanying Python code is available \footnote{\href{https://github.com/a-carson/resampling_neural_afx}{https://github.com/a-carson/resampling\_neural\_afx}}.

\section{Audio Recurrent Neural Networks}\label{sec:rnns}
Consider a continuous-time input audio signal $x(t)$ that has been sampled at a rate of $F_s=1/T$ to give $x_n$ where $n$ is an integer sample index. In this work we consider recurrent neural networks (RNNs) of the form:
\begin{subequations}\label{eq:rnn_and_fc}
\begin{align}
    \mathbf{h}_n &= f\left(\mathbf{h}_{n-1}, x_n \right) \label{eq:rnn} \\ 
    y_n   &= g \left( \mathbf{h}_n, x_n \right),\label{eq:fc}
\end{align}
\end{subequations}
where $\mathbf{h}_n \in \mathbb{R}^{S \times 1}$ is the hidden state of length $S$ and $y_n \in {\mathbb R}$ is the output signal. This class of model has been extensively used in recent years for modelling guitar amplifiers and effects pedals \cite{Wright2019RNN, Wright2020, juvela2023end, cassidy2023perceptual}. In this work we consider $f$ to be a long short-term memory (LSTM) cell, and $g$ an affine transformation (which is by definition sample rate independent).
Given a target audio signal $s_n$ obtained by processing $x$ through a specific analog device and sampling at $F_s$, the RNN in \eqref{eq:rnn_and_fc} can be trained by minimizing a loss function $\mathcal{L}(y, s)$, typically the error-to-signal ratio \cite{wright2020_perceptual} or multi-resolution spectral loss \cite{YamatotoMRSL2020, steinmetz2022}. Training in this manner gives an RNN which is optimized to the sample rate $F_s$. 
Consider now the pre-trained RNN placed in an environment where the input signal, $x'_n$, is sampled at a new rate $F_s' = L/M \times F_s$ for mutually prime non-negative integers $L$ and $M$. Here we consider and compare two methods for processing $x'_n$: the sample rate independent RNN model adjustment method, and a resampling method. 

\begin{figure} [t!]
    \centering
    \includegraphics[width=1.0 \linewidth, clip, trim={0cm, 0cm, 0cm, 0cm}]{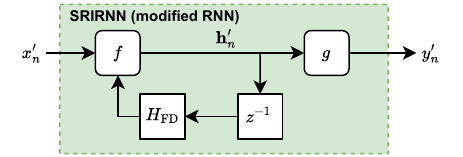}
    \caption{Model adjustment method \cite{Chowdhury2022, Carson2024, Carson2025} for RNN processing when the input signal sample rate differs from the model rate by a fractional factor of $L/M$. The fractional delay filter, $H_{\rm FD}$, implements a delay of $L/M-1$ samples.}
    \label{fig:srirnn_block}
\end{figure}

\subsection{Sample Rate Independent RNNs}\label{sec:srirnns}
The model adjustment method involves modifying the delay length (in samples) of the RNN feedback path such that the same delay duration in seconds is maintained regardless of the operating sample rate \cite{Chowdhury2022, Carson2024}. The modified RNN operating at $F'_s$, as shown in Fig. \ref{fig:srirnn_block}, can be defined as
\begin{subequations}\label{eq:srirnn}
    \begin{align}
        {\bf h}'_n &= f\left( {\bf h}'_{n-1 - \Delta}, x'_n \right)  \label{eq:srirnn_ideal} \\
        y'_n &= g\left( {\bf h}'_n, x'_n \right),  \label{eq:srirnn_g} 
    \end{align}
\end{subequations}
where $\Delta = L/M - 1$ is the delay-line length adjustment in samples \cite{Chowdhury2022, Carson2024}. For integer oversampling, i.e. $L/M - 1\in \mathbb{Z}^{+}$, and this delay-line adjustment can be implemented with a pure delay and has been shown to preserve the original RNN behaviour to a high degree of fidelity whilst having the additional benefit of reducing aliasing \cite{Carson2024}. For non-integer conversion ratios, the state at non-integer time step $n-1-\Delta$ can be approximated with a fractional delay FIR filter:
\begin{align}
    {\bf h}'_{n-1 - \Delta} \approx \sum_{k=0}^{K} l_k {\bf h}'_{n-1-k}, \label{eq:conv} 
\end{align}
where $K$ is the filter order and $l_k$ are the filter coefficients. 

\begin{figure}[t!]
    \centering
    \includegraphics[width=1.0\linewidth, trim={10, 10, 10, 10}]{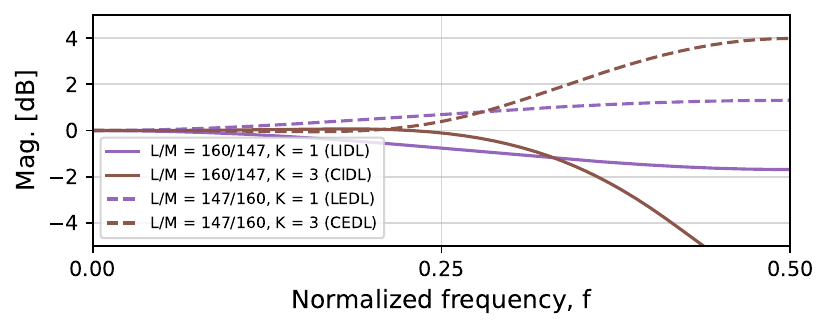}
    \caption{Magnitude response of the fractional delay filters considered for use in the model adjustment method.}
    \label{fig:lidlcidlledlcedl}
\end{figure}

Previous work \cite{Carson2025} considered Lagrange interpolation filters \cite{Laakso1996, Valimaki2000, Hermanowicz1992} and low-order optimal filters (in the mini-max sense) \cite{Putnam1997} for this task. It was found that the best choice of filter design method and order depended strongly on the weights of the original RNN model. Certain combinations of filter and RNN can lead to instability within the RNN's state-space causing catastrophic failure. These problems were especially prevalent for $L/M < 1$, where fractional extrapolation (as opposed to interpolation) is required. This extrapolation acts as a high shelving filter, thus many frequencies experience a gain greater than unity in the RNN feedback loop. In this work we consider first and third order Lagrange FIR filters as baselines. First order Lagrange filtering (for interpolation or extrapolation) was considered for its simplicity, and third order was chosen as this was shown in previous work to give the best results for both oversampling and undersampling \cite{Carson2024, Carson2025}. For $L/M > 1$ (fractional oversampling) we refer to these as linearly interpolated delay line (LIDL) and cubic interpolated delay line (CIDL) methods respectively. For $L/M < 1$ (fractional undersampling) we refer to these as linearly extrapolated delay line (LEDL) and cubic extrapolated delay line (CEDL) methods respectively. Fig.~\ref{fig:lidlcidlledlcedl} shows the magnitude response of the interpolation/extrapolation filters for $L/M = (160/147, 147/160)$. The first order all-pass filter used in \cite{Carson2024} for fractional oversampling was not considered here, as in some cases poor results were reported due to numerical errors in pole-zero cancellation. Furthermore, this method is not suitable for approximating a fractional signal advance as required here for $L/M < 1$.

Given a state size of $S$, the computational cost of the SRIRNN methods is $(K+1)S$ multiplications and $K S$ additions per sample at the operating rate $F'_s$. This does not include the operations within the RNN cell itself, which would be $L/M$ times the original number of operations per sample at the training rate $F_s$. There is no latency associated with these methods because the interpolation is carried out over previous states of the model only.

\section{Resampling Method}\label{sec:resampling}
\begin{figure}
    \centering
    \includegraphics[width=1.0 \linewidth, clip, trim={0cm, 0cm, 0cm, 0cm}]{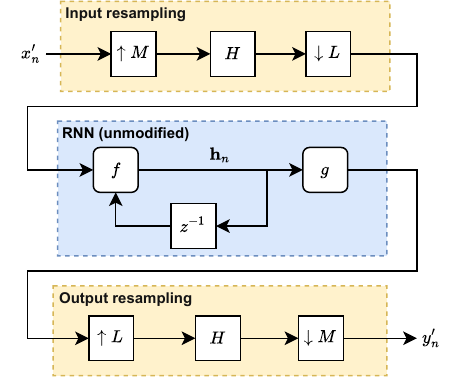}
    \caption{Resampling method investigated in this work for RNN processing when the input signal sample rate differs from the model rate by a fractional factor of $L/M$.}
    \label{fig:resampling_block}
\end{figure}
The alternative method of handling the discrepancy between the model sample rate and input signal rate involves resampling the input and output signals of the RNN such that processing occurs at the originally intended rate. This is illustrated in Fig.~\ref{fig:resampling_block} and can be defined:
\begin{subequations}\label{eq:resampling_rnn}
    \begin{align}
        {\bf h}_n &= f\left( {\bf h}_{n-1}, \mathcal{R}_{M/L}(x'_n) \right)   \\
        y'_n &= \mathcal{R}_{L/M} \Bigg( g\left( {\bf h}_n, \mathcal{R}_{M/L}(x'_n) \right)  \Bigg),
    \end{align}
\end{subequations}
where $\mathcal{R}_{L/M}(\cdot)$ denotes a sample rate conversion by a factor of $L/M$. In an offline scenario, this has become easy and can be achieved with exact FFT-based resampling  \cite{Valimaki2023}. Here, however, we investigate in detail the best resampling methods suitable for real-time use. 
In the time domain, sample rate conversion by rational fraction $L/M$ can be implemented by first expanding the signal to rate $L$, low-pass filtering, then decimating by a factor of $M$ \cite{CrochiereRabiner1981, Vaidyanathan_TB}. Note that in \eqref{eq:resampling_rnn} this means the low-pass filtering for all the resampling operations occurs at a rate of $LF_s \equiv M F'_s$. The ideal low-pass filter for this task has the magnitude response:
\begin{equation}\label{eq:lpf_ideal}
    H_{\rm  ideal}(e^{j \omega}) = \begin{cases}
        1 & |\omega| < 2 \pi f_c \\
        0 & {\rm otherwise,}
    \end{cases}
\end{equation}
where the normalized cut-off frequency, $f_c$, is set to the Nyquist limit of either $F_s$ or $F'_s$ -- whichever is lower -- to avoid spectral imaging due to expansion or aliasing due to decimation:
\begin{equation}
    f_{c, {\rm ideal}} = \frac{1}{2 \times {\rm max}(L, M)}.
\end{equation}
For the system in \eqref{eq:resampling_rnn}, the ideal cut-off frequency is therefore the same for both resampling stages $\mathcal{R}_{L/M}$ and $\mathcal{R}_{M/L}$. For this reason, in this work we will use the same filter design for both pre-RNN and post-RNN resampling.

\begin{figure}[t!]
    \centering
    \includegraphics[width=1.0 \linewidth, clip, trim={0cm, 0cm, 0cm, 0cm}]{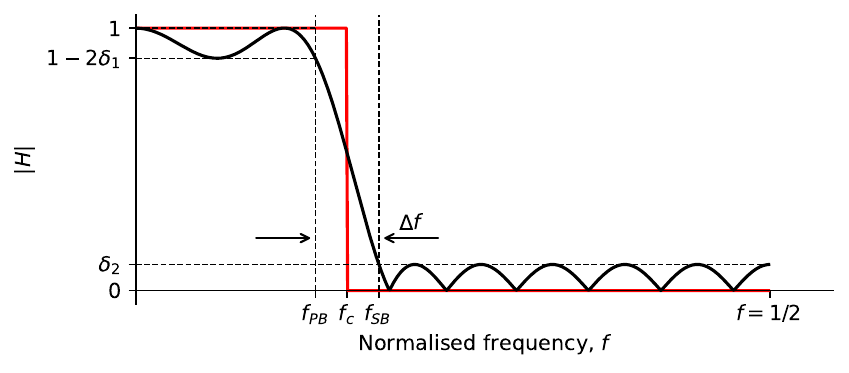}
    \caption{Magnitude response of an ideal low-pass filter (red) and an example design (black) for resampling where ${\rm max}(L, M) = 3$, $\Delta f = 1/20$, $A_p = $ \SI{1}{\dB} and $A_s = $ \SI{20}{\dB}.}
    \label{fig:filter_sketch}
\end{figure}

The filter \eqref{eq:lpf_ideal} is not practically realizable in the time domain so here we consider filter designs based on the following \textit{design variables}:
\begin{itemize}
    \item $f_{\rm PB}$: passband edge frequency (normalised);
    \item $f_{\rm SB}$: stopband edge frequency (normalised);
    \item $A_p$: maximum peak-to-peak passband ripple in \SI{}{\dB};
    \item $A_s$: minimum stopband attenuation in \SI{}{\dB}. 
\end{itemize}
From these variables we can derive:
\begin{itemize}
    \item the transition bandwidth, $\Delta f = f_{\rm SB} - f_{\rm PB}$, in which there are no requirements for attenuation;
    \item the passband ripple \textit{amplitude}, $\delta_1 = \frac{10^{A_p/20} - 1}{10^{A_p/20} + 1}$;
    \item the stopband ripple height $\delta_2 = 10^{-A_s/20}$.
\end{itemize}
Fig. \ref{fig:filter_sketch} illustrates an example filter designed from these variables. Given that the filter operates at the rate $LF_s \equiv MF'_s$, the passband and stopband edge frequencies can also be defined in Hz: $F_{\rm PB} = LF_sf_{\rm PB}$ and $F_{\rm SB} = LF_sf_{\rm SB}$ respectively.
In this work we treat \SI{44.1}{\kHz} as the base sample rate and consider sample rate conversions between this rate and higher, such that $\min(F_s, F'_s) = $ \SI{44.1}{\kHz}. Let us establish the following absolute \textit{design specifications}:
\begin{itemize}
    \item The passband is the range $(0, 16)$ \SI{}{\kHz} and the maximum deviation from unity gain within this range should be $\pm$\SI{0.5}{\dB}. This range was chosen as human hearing sensitivity drops significantly above this level \cite{Kahles2019}, and in guitar processing these frequencies would typically be attenuated by the speaker cabinet (or emulation thereof).
    \item The stopband begins at \SI{28.1}{\kHz} and the minimum stopband attenuation should be approximately \SI{120}{\dB}. This ensures that when decimating to \SI{44.1}{\kHz} any aliasing within the passband is attenuated below this threshold.
    \item The filter must be normalised to give unity gain at \SI{0}{\Hz}.
\end{itemize}
Note that in some cases, we may set the \textit{design variables} to differ from the \textit{design specifications} so long as the resultant filter satisfies these specifications, e.g. $F_{\rm PB}$ may not always be set to \SI{16}{\kHz}. We seek to design filters which give a similar or better performance to the model adjustment method according to the metrics in Sec. \ref{sec:metrics}, and to compare the associated computational cost of each method. 

\subsection{FIR Filters}
Finite impulse response (FIR) filters offer the simplest design of low-pass filter for resampling. A linear-phase FIR design involves windowing the impulse response of the ideal filter in \eqref{eq:lpf_ideal} to $N+1$ coefficients: 
\begin{equation}\label{eq:fir_win}
    h(n) = f_c {\rm sinc}\left(f_c n  \right)  w(n),
\end{equation}
where $w(n)$ is an even-symmetric window function that is zero for $|n| > N/2$, $N$ is the filter order (assumed even), and $f_c = (f_{\rm PB} + f_{\rm SB})/2$ \cite{LyonsUnderstandingDSP}. For a given order, the window function determines the trade-off between passband and stopband error \cite{Harris78_windows}. Here we use a Kaiser window, as this enables the required filter order, $N$, and window parameter $\beta$ to be determined from a specified transition bandwidth and stopband attenuation. The filter order is given by:
\begin{equation}\label{eq:kaiser_ord}
    N_{\rm kaiser} \approx \frac{A_s - 7.95}{14.36 \Delta f},
\end{equation}
and $\beta = 0.1102(A_s - 8.7)$ for $A_s > $ \SI{50}{\dB}. In this work we round up $N_{\rm kaiser}$ to the nearest even integer to ensure the group delay is an integer number of samples. Note that the Kaiser design does not allow explicit control over passband ripple.

We also consider FIR filters designed using the Parks-McClellan method (also known as the Remez exchange algorithm) \cite{parksmcclellan1972}. This gives an equiripple filter which minimizes the maximum error for a given band. For a given stopband attenuation $A_s$ and maximum passband ripple $A_p$ the required order can be approximated with Bellanger's formula \cite{Vaidyanathan_TB}:
\begin{equation}\label{eq:bellanger}
    N_{\rm eq.} = \frac{2 \log_{10}\left(\frac{1}{10 \delta_1 \delta_2}\right)}{3 \Delta f}.
\end{equation}

FIR filters for resampling can be implemented in the time domain using a polyphase decomposition \cite{Vaidyanathan_TB}. In this case the number of multiplications per unit time (MPUs) is $(N+1)/M$ and the number of additions per unit time (APUs) is $(N+1-L)/M$. For linear-phase FIR filters the latency is $N/2$ samples -- in the context of resampling this means the latency in seconds is $N/(2LF_s)$. 

\subsection{Half-band IIR Filters}
Infinite impulse response (IIR) filters generally meet a required set of specifications with a much lower filter order than FIR filters. Methods for designing IIR filters include Butterworth, Chebyshev, and elliptic designs \cite{LyonsUnderstandingDSP}. In this work, we use an elliptic \textit{half-band} IIR filter for the special case of interpolation or decimation by a factor of 2. This introduces phase distortion, but we consider this acceptable as the phase is already modified in distortion effects processing and the requirement for linear-phase audio filters is disputed \cite{FILTERSWEB07, ValimakiEQ2016}. Half-band IIR filters can be decomposed into two poly-phase branches, each of which is processed by an all-pass filter:
\begin{equation}\label{eq:iir_poly}
    H_{\rm HB}(z) = \frac{A_0(z^2) +  z^{-1} A_1(z^2)}{2},
\end{equation}
where $A_0(z)$ and $A_1(z)$ are all-pass filters with orders $n_0$ and $n_1$ respectively. The order of $H_{HB}(z)$ is $N = 2(n_0 + n_1) + 1$. Each all-pass filter can be implemented with zero latency and only $n_i$ MPUs and $2n_i$ APUs where $n_i$ is the filter order \cite{Valenzuela1984}. For a derivation of the filter coefficients the reader is referred to \cite{Valenzuela1984} or the Python code provided. 

\subsection{Proposed Designs for 44.1\,kHz $\leftrightarrow$ 48\,kHz}\label{subsec:CD_DAT_designs}
\begin{figure}[t!]
    \centering
    \includegraphics[width=1.0\linewidth,clip, trim={0 0 0 0}]{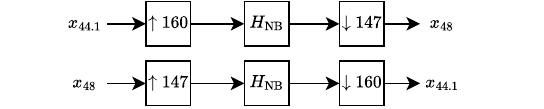}
    \caption{Single-stage resampling from \SI{44.1}{\kHz} to \SI{48}{\kHz} (top) and the reverse (bottom). $H_{\rm NB}$ is a narrow-band FIR filter. }
    \label{fig:CD_DAT_single}
\end{figure}

\begin{figure}[t!]
    \centering
    \includegraphics[width=1.0\linewidth,clip, trim={10 0 10 0}]{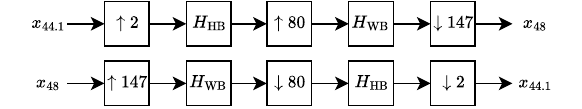}
    \caption{Two-stage resampling from \SI{44.1}{\kHz} to \SI{48}{\kHz} (top) and the reverse (bottom). $H_{\rm WB}$ is a ``wide-band'' FIR filter and $H_{\rm HB}$ is a half-band IIR filter.}
    \label{fig:CD_DAT_twostage}
\end{figure}

This sub-section outlines the proposed resampling methods for use in system \eqref{eq:resampling_rnn} when $F_s = $ \SI{44.1}{\kHz}, $L = 160$, and $M = 147$ or $F_s = $ \SI{48}{\kHz}, $L = 147$, and $M = 160$. We first consider single stage sample rate conversion as shown Fig.~\ref{fig:CD_DAT_single}, requiring the design of a narrow-band (NB) FIR filter between the expander and decimator. Furthermore, we consider decomposing the interpolation or decimation into two stages as shown Fig.~\ref{fig:CD_DAT_twostage}. Multi-stage resampling methods for this conversion ratio have been explored previously (e.g. \cite{Kashtiban_CD_DAT_2006}) but to our knowledge not in the context of RNN audio effects processing. For interpolation, our design involves first doubling the sample rate with a half-band (HB) interpolator followed by sample-rate conversion by a factor of $80/147$. The benefit of this design is the reduction in latency: the first stage can be efficiently implemented with a half-band IIR filter; and the second stage FIR filter can have a much wider transition bandwidth and therefore lower order (these wider band FIR filters are labelled WB). 

For decimation, the process is reversed: interpolate by a factor of $147/80$ then decimate by a factor of two. The list below summarises the four designs considered, Fig.~\ref{fig:CD_DAT_mag_response} shows their respective magnitude responses and Table \ref{tab:CD_DAT_filters} shows the computational costs and latency. Compared to the single-stage method, NB, the two-stage method (WB cascaded with HB) requires a similar number of operations per input sample but the latency is over 70\% less. In practice, the latency of all methods included in Table \ref{tab:CD_DAT_filters} are small enough for real-time audio processing. Chafe et al.~\cite{Chafe2010} have reported that musicians require less than 25\,ms of delay between each other for effortless interaction. A recent study \cite{Tomasetti2023} revealed that the lowest latency appearing in real-time spatial audio software plugins is 0.46\,ms but varies between 1 and 40\,ms in different products. The latency of all the methods included in this study is smaller than that (see  Table \ref{tab:CD_DAT_filters}).

\subsubsection{NB-Kaiser}
A single-stage design that meets the required specifications can be achieved with a narrow-band FIR filter using the Kaiser method.
Here we use a passband edge of $F_{\rm PB} = $ \SI{11.5}{\kHz} and stopband edge $F_{\rm SB} = $ \SI{28.1}{\kHz}, giving a normalised transition bandwidth of $\Delta f \approx 0.0024$. For $A_s = $ \SI{120}{\dB} the required filter order is $N=3318$ (calculated from  \eqref{eq:kaiser_ord} and rounded to the next even integer).
 
 \subsubsection{NB-Remez}
 An equiripple filter design was also considered for single-stage resampling. The passband and stopband edges were set to $F_{\rm PB} = $ \SI{16}{\kHz} and  $F_{\rm SB} = $ \SI{28.1}{\kHz} respectively. The maximum peak-to-peak ripple and stopband attenuation were set to $A_s = $ \SI{0.5}{\dB} and $A_s = $ \SI{120}{\dB}. The required order was estimated to be $N = 2544$ [from \eqref{eq:bellanger}], and the filter coefficients were obtained using the Remez exchange algorithm with a stopband weighting of $\delta_1 / \delta_2$.

\subsubsection{HB-IIR + WB-Kaiser}
\begin{figure*}[h!!]
    \centering
    \includegraphics[width=1.0\linewidth]{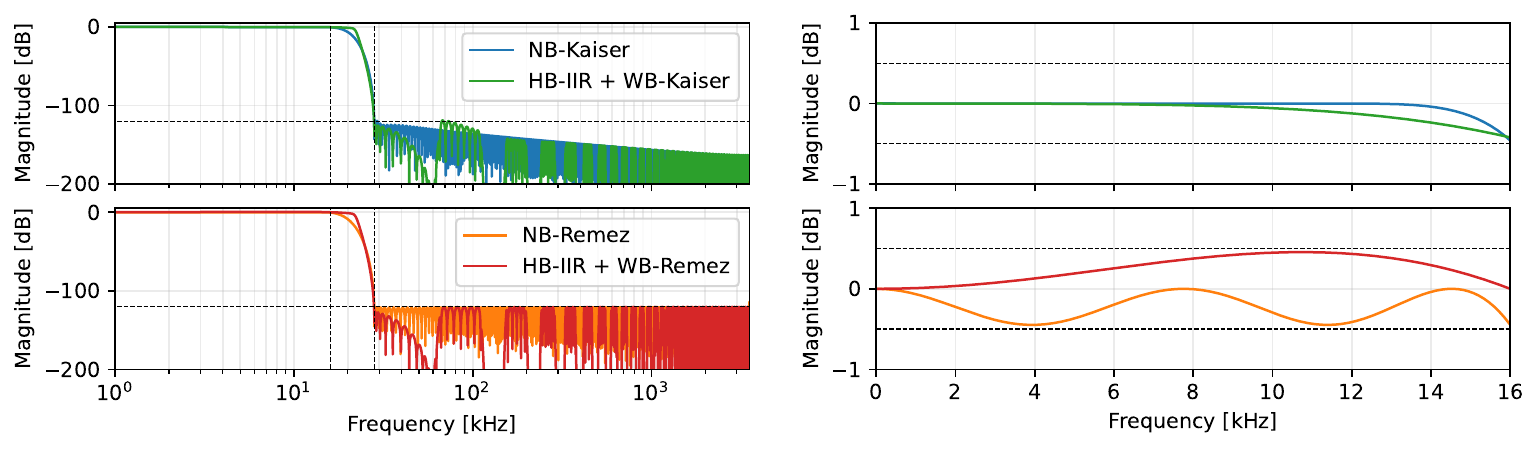}
    \caption{(left) Magnitude response above \SI{1}{\kHz} and (right) passband details of the filters used for resampling between \SI{44.1}{\kHz} and \SI{48}{\kHz}. On the left, the dashed lines indicate the passband and stopband edges and the stopband ripple specification. On the right, the dashed lines are the $\pm$0.5-dB passband ripple tolerances. All filters fulfill the specifications sufficiently but there are significant differences in computational costs, see Table \ref{tab:CD_DAT_filters}.}
    \label{fig:CD_DAT_mag_response}
\end{figure*}

The half-band filter used in the two-stage methods was designed with a stopband edge of \SI{28.1}{\kHz} (and therefore a passband edge of \SI{16}{\kHz} due to symmetry). A power-symmetric elliptic filter of order 13 gives a stopband attenuation \SI{119.7}{\dB} which was deemed close enough to the specifications. The power symmetry allows the filter to be decomposed into two all-pass filters as in \eqref{eq:iir_poly} with orders $n_0 = n_1 = 3$. The FIR filter for interpolation by $147/80$ or decimation by $80/147$ can then be designed with a stopband edge of $88.2 - 28.1$ = \SI{60.1}{\kHz}. For the Kaiser windowed FIR design the passband edge was set to \SI{0}{\Hz} whilst still ensuring a maximum attenuation of \SI{0.5}{\dB} below \SI{16}{\kHz}, giving an order of $N=916$. The cascaded magnitude response of the half-band IIR filter and FIR filter is shown in Fig.~\ref{fig:CD_DAT_mag_response}.

\subsubsection{HB-IIR + WB-Remez}
The FIR filter for the multi-stage Remez method was designed using the exact same method as in the single-stage Remez design but with the higher stopband edge of \SI{60.1}{\kHz}, giving a filter order $N=698$. The same half-band interpolator/decimator as above was used. 

\begin{table}[]
    \centering
     \caption{Computational costs for a single instance of the proposed filter designs used in resampling between \SI{44.1}{\kHz} and \SI{48}{\kHz}. The ``WB'' designs are cascaded with the HB-IIR in practice, as shown in Fig.~\ref{fig:CD_DAT_twostage}, so the relevant rows must be summed for the total computational load and latency. }
     \resizebox{\linewidth}{!}{
    \begin{tabular}{lrrrrrr}
        \toprule
        \multirow{2}{*}{\bf Design} & \bf{Order}, & \multicolumn{3}{c}{\bf{Operations$^*$}} & \multicolumn{2}{c}{\bf{Latency}}\\
         & $N$ & MPUs & APUs & Total & Samples & ms \\
        \midrule
        NB-Kaiser & 3318 & 22.58 & 21.49 & 44.07 & 1659 & 0.235 \\
        NB-Remez & 2254 & 17.38 & 16.29 & 33.67 & 1277 & 0.181 \\
        WB-Kaiser & 916 & 12.48 & 11.39 & 23.87 & 458 & 0.065 \\
        WB-Remez & 698 & 9.51 & 8.42 & 17.93 & 349 & 0.049 \\
        HB-IIR & 13 & 6 & 12 & 18 & 0 & 0 \\
        HB-FIR$^\dagger$ & 54 & 14 & 28 & 42 & 27 & 0.306 \\
        \hline
        \multicolumn{7}{l}{$^*$per sample at $F_s = $ \SI{44.1}{\kHz}.}\\
        \multicolumn{7}{l}{$^\dagger$ used in Sec. \ref{sec:oversampling}.}
    \end{tabular}}
    \label{tab:CD_DAT_filters}
\end{table}

\section{Test Signals and Metrics}\label{sec:metrics}
In all experiments the input signals to the models were sine tones with $f_0$ ranging from \SI{27.5}{\Hz} to \SI{4186}{\Hz}. These were generated at both the original model rate, $F_s$ and the different rate of $F'_s$:
    \begin{align}
        x_n =  g \sin (2 \pi f_0 n/F_s),\qquad
        x'_n = g \sin (2 \pi f_0 n/F'_s), 
    \end{align}
where $g$ is a constant gain term. For each $f_0$, the ground truth $y$ was generated by passing $x$ through the original model architecture in \eqref{eq:rnn_and_fc}. The output signal under examination, $y'$, is generated by processing $x'$ through either the model adjustment method \eqref{eq:srirnn} or resampling method \eqref{eq:resampling_rnn}. To enable separate analysis of the (desired) harmonic distortion components and the aliased components of the output signals, we construct alias-free bandlimited versions using the following method. First, $y$ and $y'$ are windowed with a Chebyshev window with 120\,dB stopband attenuation and the discrete Fourier transform taken to obtain the spectra $Y$ and $Y'$. The amplitudes and phases of the harmonic components are extracted from each spectrum and used to synthesise ideal alias-free versions of the signals $y_{\rm BL}$ and $y'_{\rm BL}$. These bandlimited signals are both synthesised at the original model rate, $F_s$. 

Four metrics are then considered: the error-to-signal ratio (ESR), magnitude-error-to-signal ratio (MESR), aliasing-to-signal ratio (ASR), and noise-to-mask ratio (NMR). In all four metrics, a lower result is better. Here we use the terms ``error'' to describe errors in harmonic signal components and ``noise'' for aliasing or more generally non-harmonic error.

The ESR is defined as the energy ratio between the time-domain error and the ground truth signal: 
\begin{equation}
    {\rm ESR} = \frac {\left \lVert y_{\rm BL} - y'_{\rm BL} \right \rVert_2^2}
    {\left \lVert y_{\rm BL} \right \rVert_2^2},
\end{equation}
where $\left \lVert \cdot \right \rVert_2$ is the L-2 norm. Note this is typically used as a loss function in RNN training \cite{Wright2020}. Here we use the bandlimited versions of the signals to ignore aliasing noise, since these are measured by the ASR and NMR metrics. The ESR is sensitive to phase errors, which can lead to poor correlation with perception (see e.g. \cite{YingShou2024}). Therefore we also consider the MESR, defined as the error ratio between the \textit{magnitude} error and the ground truth signal:
\begin{eqnarray}
    {\rm MESR} = \frac{\lVert |Y_{\rm BL}| - |Y'_{\rm BL}| \rVert_2^2}{\lVert Y_{\rm BL} \rVert_2^2}.
\end{eqnarray}
where $|\cdot|$ is the complex modulus. $Y_{\rm BL}$ and $Y'_{\rm BL}$ are the spectra of $y_{\rm BL}$ and $y'_{\rm BL}$ respectively. The MESR can be considered a measure of the colouration introduced by the methods, the less of which is better.

Time-domain resampling will inevitably introduce some amount of aliasing, both when interpolating (through spectral imaging) and decimating, because an ideal filter (Eq. \eqref{eq:lpf_ideal}) cannot be realised in practice \cite{Vaidyanathan_TB}. Furthermore, the nonlinear activation functions within the RNNs themselves generate aliasing alongside the desired harmonic distortion \cite{Wright2020, Vanhatalo2024}. The simplest measure of aliasing in spectrum $Y$ is the ASR, defined as the energy ratio between aliasing noise and the desired harmonic components:
\begin{eqnarray}
    {\rm ASR} = \frac{\lVert Y_{\rm BL} - Y \rVert_2^2}{\lVert Y_{\rm BL} \rVert_2^2},
\end{eqnarray}
where $Y_{\rm BL}$ is the bandlimited alias-free spectrum. 

For a more perceptually-informed measure of aliasing, we consider the NMR, which was originally developed for evaluating perceptual audio coding \cite{NMR_ITU-R}. It is defined as the energy ratio between non-harmonic (noise) components and the simplified masking threshold of desired harmonic components, and it uses a human ear filter model to account for our frequency-dependent hearing sensitivity  \cite{NMR_ITU-R, KabalNMR2003}. The NMR has been used as a measure of aliasing in other related works\cite{Kahles2019, Lehtonen_2012_saw}. Here we use the MATLAB implementation provided by Zheleznov and Bilbao \cite{zhelevnov24}.

Given that RNNs are nonlinear, the choice of input gain $g$ will affect the results of all the above metrics. 
In our experiments, we use crowd-sourced RNN models as case studies, and because we do not have access to the original training data we cannot estimate the level (gain) of the signals used to train the models (which will be different for every model).  This makes choosing $g$ for our experiments somewhat arbitrary -- here we simply choose $g=0.1$ for all experiments. We acknowledge this will affect some numerical results, but deem the experiments across $f_0$ and between many models as sufficient for drawing conclusions on the properties of the investigated methods.
\section{Experiment: SRIRNN vs Resampling}\label{sec:CD_DAT}
This section applies and compares the SRIRNN method (Sec. \ref{sec:srirnns}) and the resampling method (Sec. \ref{sec:resampling}) on RNN models of guitar distortion effects. We focus on two opposite cases: $(F_s, F'_s) = (44.1, 48)$ kHz and $(F_s, F'_s) = (48, 44.1)$ kHz. The four resampling filter designs in Sec. \ref{subsec:CD_DAT_designs} are compared alongside first order (LIDL/LEDL) and third order (CIDL/CEDL) baseline methods. The results for the ``naive'' method are also presented---these are obtained by processing $x'$ through the original RNN despite the sample rate discrepancy. The ASR and NMR of the original, unmodified model output are provided in the results as a reference---obtained by passing $x$ through the original model when there is no sample rate discrepancy ($L=M=1$). The computational costs and latencies of the compared methods are shown in Table \ref{tab:compute_costs}. The internal of operations of the RNN cell are not included in the table. In the SRIRNN method, the RNN cell operates at the input signal rate, $F'_s$, whereas in the resampling method the RNN cell operates at the original training rate $F_s$. This means that for $F'_s > F_s$, the resampling method makes the RNN cell cheaper to operate, but for $F_s > F'_s$ it is more expensive. For discrepancies between \SI{44.1}{\kHz} and \SI{48}{\kHz}, the difference in RNN cell cost will be approximately $\pm 8$ \%. In most cases, it is likely that this will outweigh the difference in filtering operations listed in Table~\ref{tab:compute_costs}, but this will depend on RNN size, implementation and hardware. These factors should be considered when determining which method is cheaper overall.

\begin{table}[]
    \centering
    \caption{Computational cost of the investigated methods for model/input rate discrepancies between \SI{44.1}{\kHz} and \SI{48}{\kHz} and an RNN with $S=80$ states.}
    \begin{tabular}{cccc}
        \toprule
        $L/M$ & Method & Filter ops$^\dagger$ & Latency (ms)\\
        \midrule
        \multirow{4}{*}{160/147$^*$} & NB-Kaiser  & 88.14 & 0.47\\ 
        & NB-Remez    & 67.35 & 0.36\\ 
        & HB-IIR + WB-Kaiser  & 83.73 & 0.13\\ 
        & HB-IIR + WB-Remez  & 71.86 & 0.10 \\
        \midrule
         \multirow{2}{*}{160/147} & LIDL           & 261.22 & 0 \\ 
        & CIDL        & 609.52 & 0 \\ 
        \midrule
        \multirow{2}{*}{147/160} & LEDL            & 240 & 0 \\ 
        & CEDL     & 560 & 0 \\ 
        \bottomrule
        \multicolumn{4}{l}{$^\dagger$ operations per sample at $F_s = $ \SI{44.1}{\kHz}.}\\
        \multicolumn{4}{l}{$^*$ same cost for $L/M = 147/160$}\\
    \end{tabular}
    \label{tab:compute_costs}
\end{table}

\subsection{Models Trained at \SI{44.1}{\kHz}; Inference at \SI{48}{\kHz}.}
\begin{figure}[t!]
    \centering
    \includegraphics[width=1.0\linewidth]{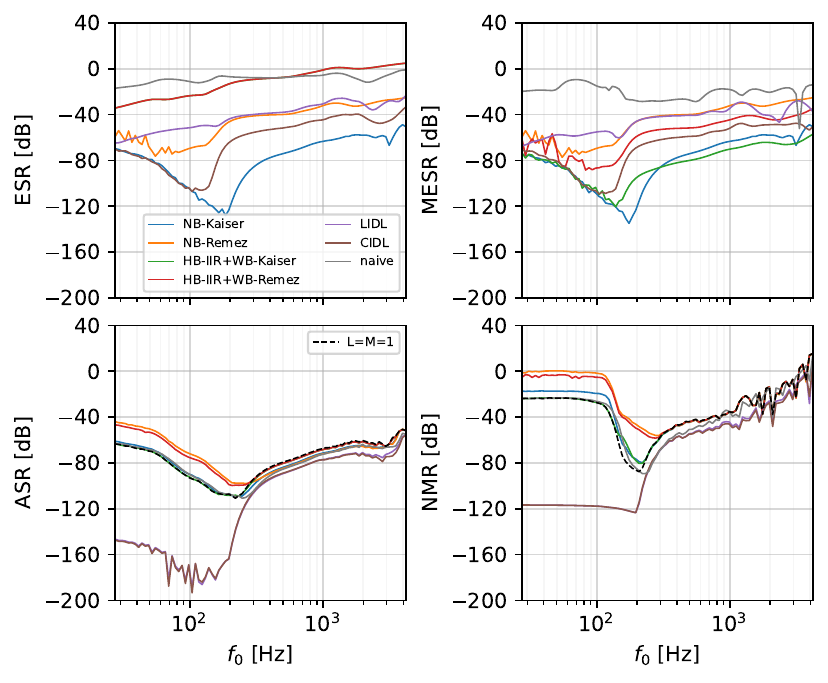}
    \caption{Output signal metrics for the \texttt{RockmanXPR\_HighGain} model trained at $F_s = $ \SI{44.1}{\kHz} and with $L=160$, $M=147$ at inference. Lower values are better. Black dashed lines show the aliasing metrics in the original system ($L=M=1$).}
    \label{fig:Rockman_metrics_CD_DAT}
\end{figure}

\begin{figure}[t!]
    \centering
    \includegraphics[width=1.0\linewidth]{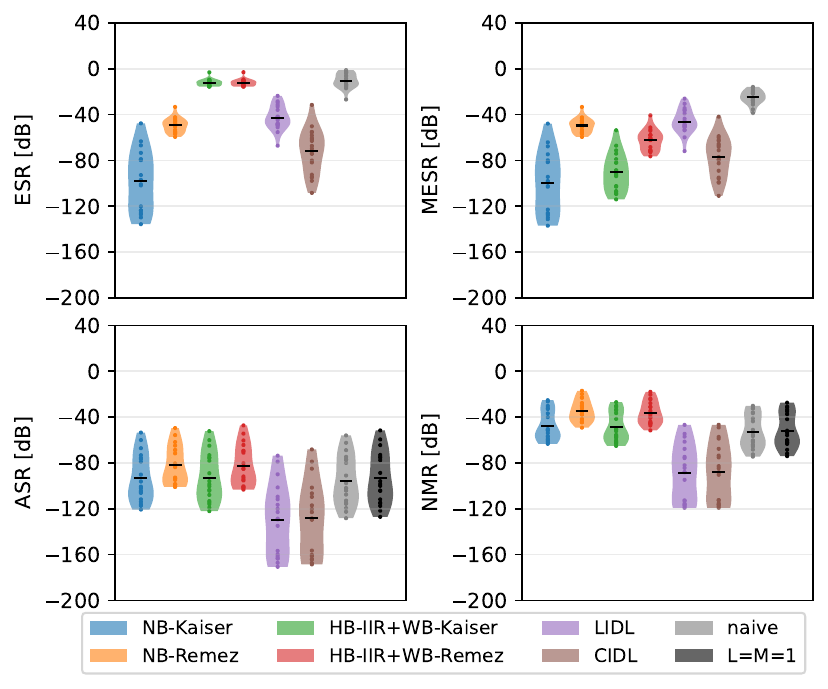}
    \caption{Output signal metrics averaged across input tone frequency for 19 LSTM models trained at $F_s = $ \SI{44.1}{\kHz} with $L=160$, $M=147$ at inference. Each dot represents a different model with some dots overlapping. Aliasing metrics for the original system are shown in black ($L=M=1$).}
    \label{fig:violin_metrics}
\end{figure}

For a set of models trained at \SI{44.1}{\kHz} we use the GuitarML ToneLibrary\footnote{\href{https://guitarml.com/tonelibrary/tonelib-pro.html}{https://guitarml.com/tonelibrary/tonelib-pro.html}}. We consider one model (\texttt{RockmanXPR\_High\_Gain}) as a case study used to inform the design process and a selection of 18 considered as an unseen test set---the same set used in our previous work \cite{Carson2024}. These all share the same LSTM architecture with $S=80$ states (including hidden and cell states). Fig.~\ref{fig:Rockman_metrics_CD_DAT} shows the signal metrics against input sine tone frequency for the \texttt{RockmanXPR} model, and Fig.~\ref{fig:violin_metrics} shows the signal metrics averaged over input frequency for all remaining models.

As we would expect due to the linear-phase property of the FIR filter designs, which preserves the signal phase, the ESR is considerably lower for the single-stage resampling methods than the nonlinear phase two-stage methods. The proposed NB-Kaiser method gives the lowest ESR on average (at the cost of the most latency), with the baseline CIDL method giving the second lowest (at the cost of the most operations sample). In terms of MESR, both the NB-Kaiser and HB-IIR+WB-Kaiser methods give a similar spread of results to the baseline CIDL method with many fewer operations per sample, and only \SI{0.1}{\ms} of latency -- equivalent to that which is experienced by a listener at a distance of \SI{3.4}{\cm} from a sound source. Comparing the two FIR design methods, the equiripple designs create more error in the harmonic components but this is unsurprising due to the greater passband ripple by design. 

\begin{figure}
    \centering
    \includegraphics[width=1.0\linewidth, clip, trim={0, 0.2cm, 0, 0.2cm}]{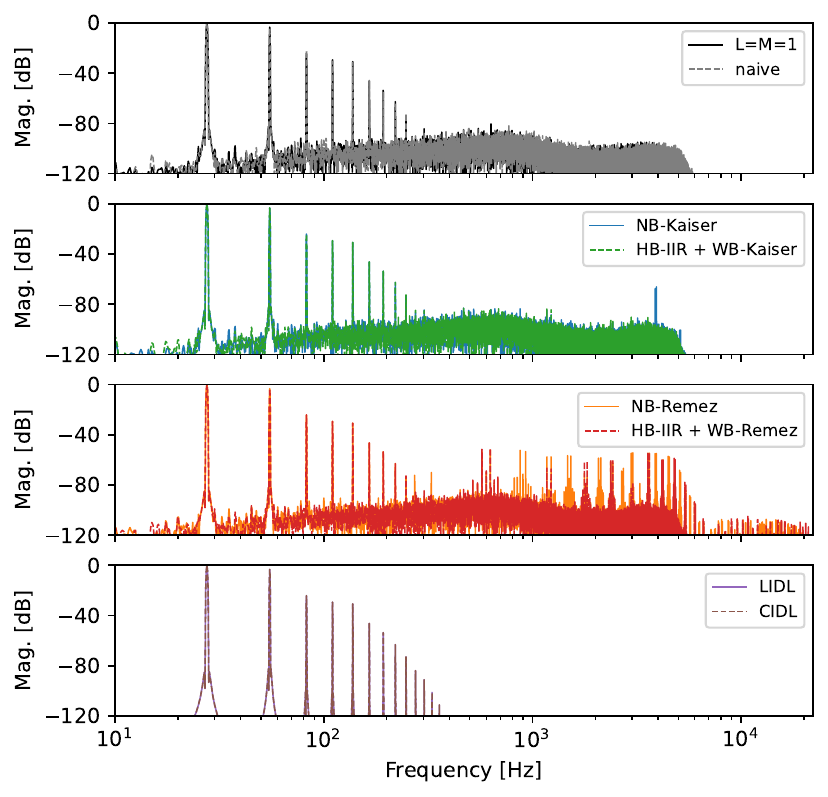}
    \caption{Spectrum of a $f_0 = $ \SI{27.5}{\Hz} tone processed through the \texttt{RockmanXPR\_High\_Gain} model trained at $F_s = $~\SI{44.1}{\kHz} and with $L=160$, $M=147$ at inference.}
    \label{fig:rockman_tones}
\end{figure}

In terms of aliasing, the Kaiser-based designs give acceptable quality with a similar ASR and NMR results to the original model (with no resampling or delay-line interpolation). The Remez designs, however, result in an increase in aliasing for low-frequency inputs to certain models. Example output spectra from the \texttt{Rockman\_XPR} case study are shown in Fig.~\ref{fig:rockman_tones}. Aliased components that are generated at the input resampling stage are amplified through the LSTM, so we conclude that a uniform stopband attenuation of \SI{120}{\dB} is not sufficient and instead the monotonically decreasing ripple height of the Kaiser window method is preferable for aliasing suppression.

The LIDL and CIDL methods generally produce the least aliasing out of all the methods, and in some cases less than the original model as shown in Fig.~\ref{fig:Rockman_metrics_CD_DAT} and Fig.~\ref{fig:rockman_tones}. The reduction in aliasing can firstly be attributed to the fact that nonlinear processing occurs at the higher rate of \SI{48}{\kHz} and therefore comes at an increased computational cost (even without considering interpolation cost). Furthermore, the LIDL and CIDL filters act as low-pass filters within the LSTM feedback loop therefore high-frequency inputs to the non-linear activations functions will be suppressed.

\subsection{Models Trained at \SI{48}{\kHz}; Inference at \SI{44.1}{\kHz}.}
For a selection of models trained at \SI{48}{\kHz} we consider models intended for the AIDA-X plug-in\footnote{\href{https://github.com/AidaDSP/AIDA-X}{https://github.com/AidaDSP/AIDA-X}} obtained from Tone3000\footnote{\href{https://tone3000.com/}{https://tone3000.com/}}. These have a variety of hidden sizes ranging from 12 to 40. For an inference input signal rate of \SI{44.1}{\kHz} we can use the same resampling filters as in the previous section but with the reverse configuration: interpolation at input stage and decimation at the output. 

\begin{figure}[t!]
    \centering
    \includegraphics[width=1.0\linewidth]{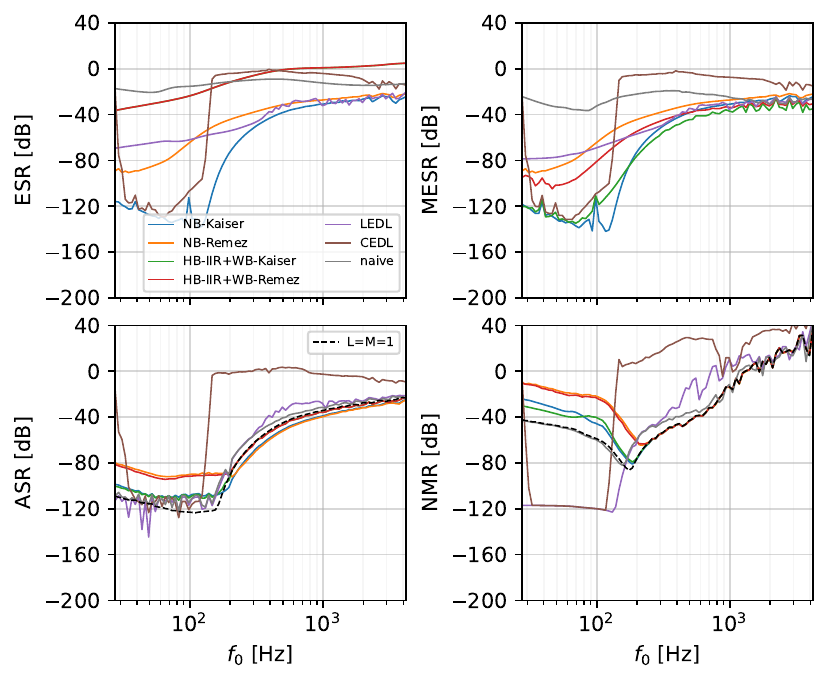}
    \caption{Output signal metrics for the \texttt{BossDs-1} model trained at $F_s = $ \SI{48}{\kHz} and with $L=147$, $M=160$ at inference. Black dashed line show the aliasing metrics in the original system ($L=M=1$).}
    \label{fig:ds1_metrics}
\end{figure}

\begin{figure}[t!]
    \centering
    \includegraphics[width=1.0\linewidth]{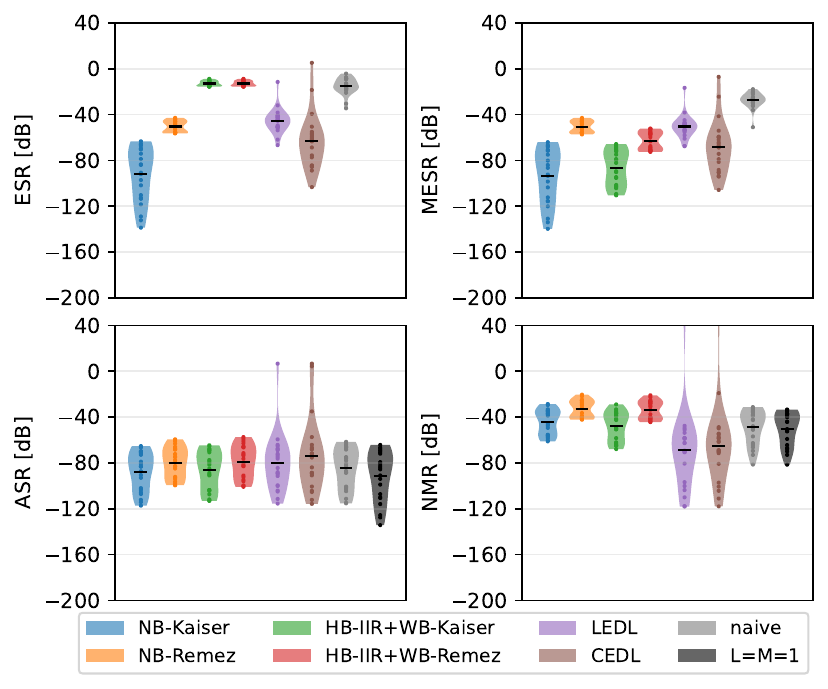}
    \caption{Output signal metrics averaged across input tone frequency for 22 LSTM models trained at $F_s = $ \SI{48}{\kHz} with $L = 147$, $M=160$ at inference. Each dot represents a different model with some dots overlapping. Aliasing metrics for the original system are shown in black ($L=M=1$).}
    \label{fig:violins_48k}
\end{figure}

Fig.~\ref{fig:ds1_metrics} shows the results from our chosen case study, the \texttt{BossDs-1(Dist\_12oclock)} model---chosen specifically because it exhibits the problematic unstable behaviour which can occur when using the SRIRNN for undersampling \cite{Carson2025}. It can be seen that for mid to high frequencies the CEDL method fails, giving poorer quality by all metrics compared to the naive method. The results for the resampling methods show a similar trend to the previous section---the equiripple designs show more aliasing for low frequency inputs compared to the Kaiser designs. 
The trends seen with the \texttt{Boss-Ds1} also appear when plotting the frequency-averaged metrics for the remaining 22 models tested, as in Fig.~\ref{fig:violins_48k}. On average, the NB-Kaiser and HB-IIR+WB-Kaiser methods give the best results in terms of MESR. Pronounced outliers can be seen in the LEDL and CEDL results, corresponding to models where the undersampling has caused the models to become unstable \cite{Carson2025}. The resampling methods avoid this instability since the model feedback loop is unaltered, resulting in no such outliers.

\section{Integer Oversampling}\label{sec:oversampling}
\begin{figure}[t!]
    \centering
    \includegraphics[width=1.0 \linewidth, clip, trim={0cm, 0cm, 0cm, 0cm}]{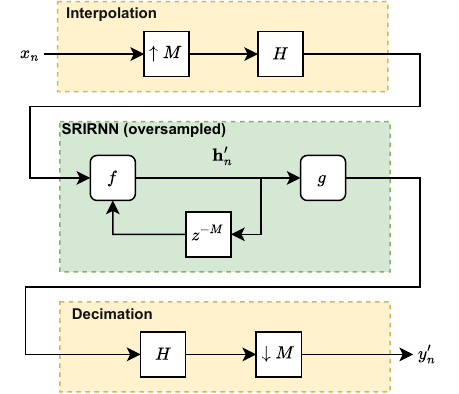}
    \caption{Combined resampling and model adjustment methods for oversampling by an integer factor $M$. The input signal is at the intended model sample rate but oversampling is used to reduce aliasing.}
    \label{fig:method_oversampling}
\end{figure}

Oversampling is used in non-linear audio processing to reduce aliasing. In our previous work we showed that the model adjustment method made this possible for RNNs \cite{Carson2024}. Here we extend this work by investigating filters for the required pre-RNN interpolation stage and post-RNN decimation stage. This system is shown in Fig.~\ref{fig:method_oversampling} and can be defined as:
\begin{subequations} \label{eq:oversampling_rnn}
    \begin{align}
        {\bf h}'_n &= f\left( {\bf h}'_{n-M}, \mathcal{R}_{M}(x_n) \right)  \\
        y_n &= \mathcal{R}_{1/M} \Big( g\left( {\bf h}'_n, \mathcal{R}_{M}(x_n) \right)\Big).\label{eq:resampling_g} 
    \end{align}
\end{subequations}
We focus on the case where the input sample rate is $F_s=$~ \SI{44.1}{\kHz} and the oversampling ratio is a power of two, $M=\{2, 4, 8\}$. Related work by Kahles et al. \cite{Kahles2019} investigated filters for oversampling by a factor of 8, using a hard-clipper as the nonlinear function. Here we consider the reported best performing method as a baseline, and propose cascaded IIR and FIR designs as described below.

\begin{figure*} [t!]
    \centering
    \includegraphics[width=1.0\linewidth]{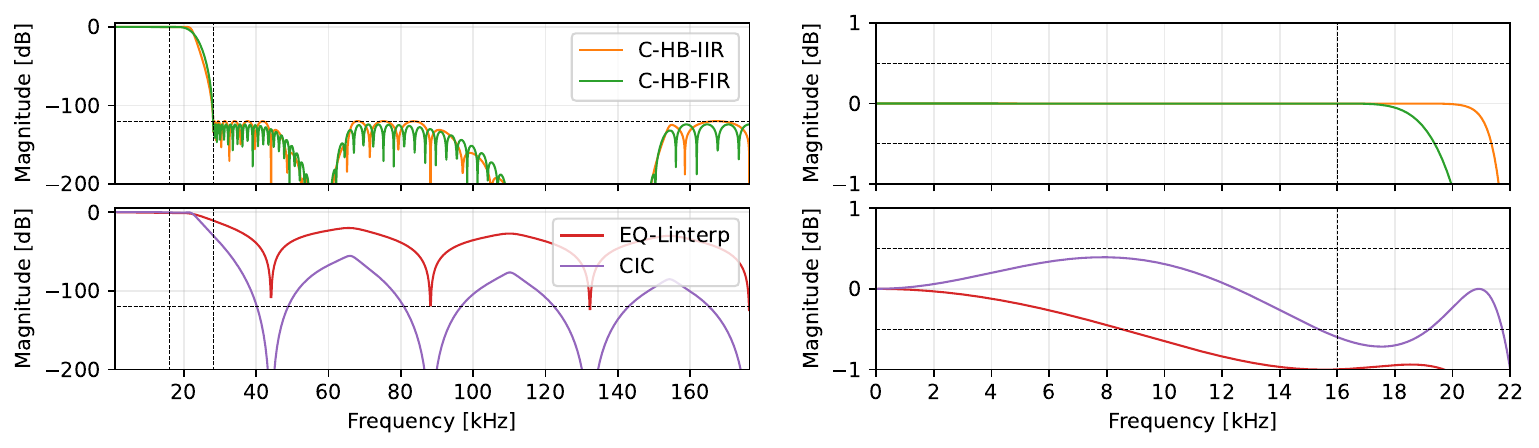}
    \caption{Interpolation and decimation filters for $M=8$ times oversampling. The dashed lines indicate the passband and stopband edges,  the stopband ripple specification (left) and the $\pm$0.5-dB passband ripple tolerances (right). The C-HB-IIR and C-HB-FIR are used for both interpolation and decimation in our proposed implementation, whereas the baseline method uses EQ-Linterp at the input and CIC for decimation at the output \cite{Kahles2019}.}
    \label{fig:M=8_filters}
\end{figure*}

\subsection{Proposed Designs}
The investigated filters for interpolation and/or decimation are enumerated below. The computational costs are reported, but note that this does not include the operations of the RNN itself, which will be $M$ times the original number of operations per sample, regardless of the method chosen. In practice, the target hardware should be considered when choosing the oversampling factor $M$, as for high $M$ the RNN may no longer be real-time capable. The magnitude response of the considered filters for $M=8$ are shown in Fig.~\ref{fig:M=8_filters}.
\subsubsection{Cascaded Half-band IIR (C-HB-IIR)}
Signal interpolation or decimation by a factor $M$ where $M$ is a power of two can be implemented by cascading $\log_2(M)$ half-band interpolators or decimators \cite{Vaidyanathan_TB}. Here we consider the same HB filter as in Sec. \ref{sec:CD_DAT}  composed of parallel all-pass filters with orders $n_0 = n_1 = 3$ giving 18 operations per input sample (for a single stage). For $M=8$ we require three cascaded filters and therefore the load is $(1 + 2 + 4) \times 18 = 126$ operations per input sample.

\subsubsection{Cascaded Half-band FIR (C-HB-FIR)}
An equivalent FIR half-band design which meets the design specifications can be designed using the Remez exchange algorithm and the method described in \cite{Vaidyanathan_HBFIR_1987}. The resulting filter has order $N=54$ but only $(N/2 + 1)/2 = 14$ non-zero unique coefficients. Exploiting this symmetry in the coefficients makes the computational cost 14 multiplications per input sample and 28 additions per input sample. For $M=8$ the load is therefore 294 operations per input sample. 

\subsubsection{Equalized linear interpolation (EQ-Linterp)}
As a baseline for the interpolation stage when $M=8$ we consider equalized linear interpolation proposed by Kahles et al.~\cite{Kahles2019}. A first-order high shelving filter (HSF) with coefficients $b_0 = 1.234$, $b_1 = 0.270$, and  $a_1 = 0.504$ is placed before the expansion stage to equalise the low-pass effect of the linear interpolation \cite{Kahles2019}. It would be possible to apply a gain factor to satisfy the passband tolerance of $\pm$\SI{0.5}{\dB} but we chose to preserve unity gain at dc to give a fairer comparison when measuring ESR. Note that this filter cascade is only used in the interpolation stage. The cascaded HSF and interpolation requires 5 multiplications and 3 additions per input sample, making it very cheap to implement.

\subsubsection{Cascaded Integrator-Comb (CIC)}
As a baseline for the decimation stage we consider the Cascaded Integrator-Comb (CIC) filter proposed by Hogenauer \cite{Hogenauer_1981_CIC} and used in \cite{Kahles2019}. This CIC filter used here is composed of a cascade of $N=6$ integrators and comb filters. The same HSF used in \cite{Kahles2019} is applied after decimation to flatten the passband. The total number of operations per input sample is 113 for $M=8$. As shown in Fig.~\ref{fig:M=8_filters}, this filter does not meet the \SI{120}{\dB} stopband attenuation criteria so we would expect this to create more aliasing than the half-band methods.

\subsubsection{FFT-based}
As a ground truth reference (or ``top-line'' method) we additionally consider exact FFT resampling \cite{Valimaki2023}. Note that this is not suitable for real-time implementation.

\subsection{Case Study: MesaMiniRec with 8x Oversampling}
\begin{figure}
    \centering
    \includegraphics[width=1.0\linewidth, clip, trim={0, 0.2cm, 0, 0.2cm}]{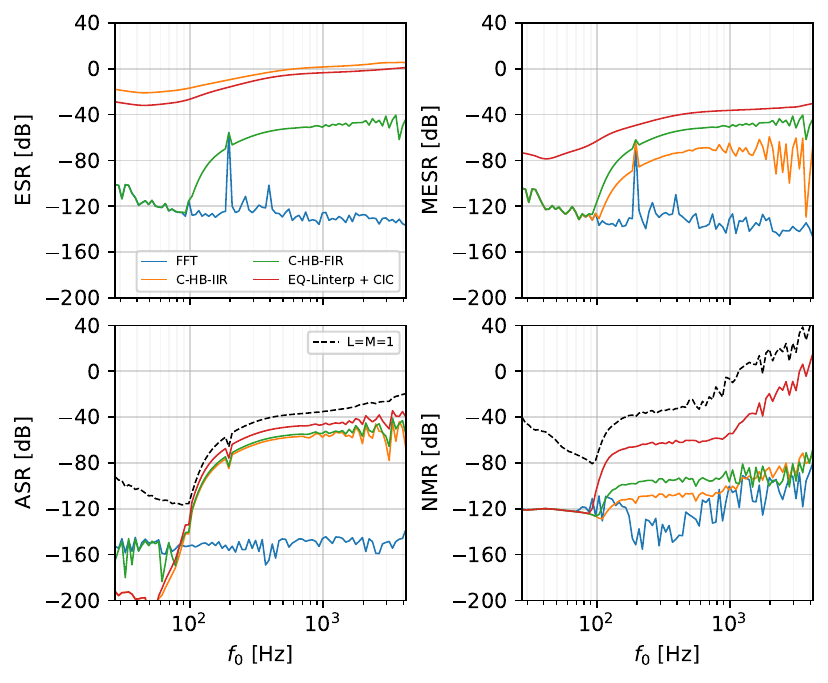}
    \caption{Comparison of interpolation and decimation filters for $M = 8$ times oversampling on the \texttt{MesaMiniRec\_HighGain} model. The black dashed line indicates the output with no oversampling.}
    \label{fig:mesa_8x_oversampling}
\end{figure}

To evaluate aliasing reduction, we consider the \texttt{MesaMiniRec\_HighGain} model as this showed the most aliasing at the base sample rate. Fig.~\ref{fig:mesa_8x_oversampling} shows the ESR, MESR, ASR, and NMR for various input sine tone frequencies with an input gain of $g=0.1$. As expected, FFT resampling at both input and output stages gives the best performance on all metrics. The time-domain resampling methods do considerably worse in terms of ASR for high frequency inputs, but the NMR results show a close match between the half-band IIR and FIR methods and the FFT method. This suggests that the aliasing measured in the ASR plot is perceptually insignificant. The baseline EQ-Linterp+CIC method \cite{Kahles2019} performs poorer in terms of NMR with the highest input frequencies exceeding the \SI{-10}{\dB} rule-of-thumb threshold of perceptibility quoted in \cite{Kahles2019, Lehtonen_2012_saw}.

\begin{figure}
    \centering
    \includegraphics[width=1.0\linewidth, clip, trim={0, 0.2cm, 0, 0.2cm}]{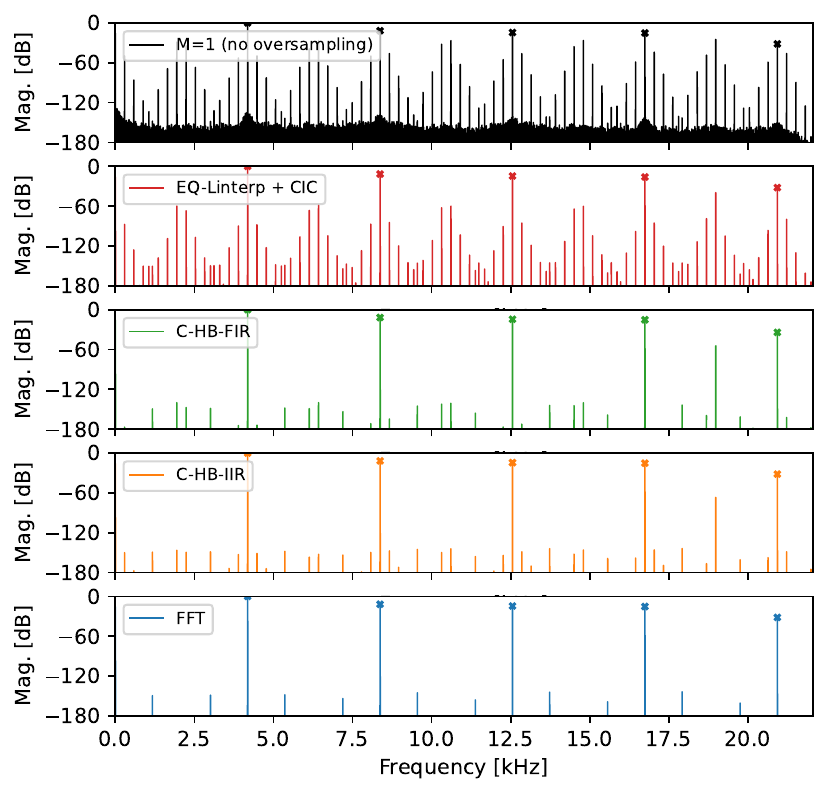}
    \caption{Spectrum of a \SI{4186}{\Hz} tone with $g=0.1$ processed through the \texttt{MesaMiniRec\_HighGain} model with $M=8$ oversampling. Crosses mark the desired harmonic components whereas the other spectral components are caused by aliasing.}
    \label{fig:mesa_tones_M=8}
\end{figure}

Fig.~\ref{fig:mesa_tones_M=8} shows the spectrum of a \SI{4186}{\kHz} sine tone (at the upper limit of tested tones) processed through the model using the different resampling filters. In the proposed half-band IIR and FIR methods all aliased components are suppressed below \SI{-150}{\dB} except the alias of the sixth harmonic which appears at $44.1 - 6 \times 4.186 = $ \SI{18.984}{\kHz} with amplitude \SI{-65}{\dB}. This is to be expected because the ideal harmonic falls within the transition band of the filters (by design), and thus does not experience \SI{120}{\dB} attenuation. We deem this acceptable behaviour as aliased components above \SI{16}{\kHz} are unlikely to be perceptible, and this explains why there is a much larger discrepancy in ASR between the FFT and HB methods compared to the perceptually-informed NMR measurement (see Fig.~\ref{fig:mesa_8x_oversampling}). 

Fig.~\ref{fig:mesa_tones_M=8} also confirms that the baseline EQ-Linterp + CIC method \cite{Kahles2019} creates visibly more aliasing than the HB methods and this is reflected in the measured NMR and ASR values in Fig.~\ref{fig:mesa_8x_oversampling}. Further investigation showed that it is the linear interpolation at the input stage that is primarily responsible for the extra aliasing -- see Table \ref{tab:cross_matrix} which shows the NMR for all combinations of the considered interpolation and decimation filters. However, even with exact FFT resampling at the input stage the CIC decimation filter creates more aliasing than the proposed half-band methods. 

\begin{table}[t!]
\centering
\caption{NMR [dB] of a \SI{4186}{\Hz} tone processed through the \texttt{MesaMiniRec\_HighGain} model for different combinations of interpolation and decimation filter, and $M=8$ times oversampling. A lower NMR is better.}
\begin{tabular}{c|cccc}
\toprule
\backslashbox{Interp.}{Decim.} & FFT   & C-HB-IIR & C-HB-FIR & CIC   \\
\hline
FFT      & -80.39  & -78.17  & -80.33 & -48.05$^{\dagger}$ \\
C-HB-IIR & -74.35  & -72.17  & -74.64 & -48.05$^{\dagger}$ \\
C-HB-FIR  & -68.3  &  -68.5  & -68.15 & -48.01$^{\dagger}$ \\
EQ-Linterp  & 15.17$^*$  &  15.17$^*$  &  15.17$^*$ &  15.23$^{\dagger}$$^*$ \\
\bottomrule
\multicolumn{5}{l}{$^{\dagger}$ worst result in each row}\\
\multicolumn{5}{l}{$^*$ worst result in each column}
\end{tabular}
\label{tab:cross_matrix}
\end{table}

\subsection{Further \texttt{HighGain} Models: 2x, 4x, and 8x Oversampling}
\begin{figure}[t]
    \centering
    \includegraphics[width=1.0\linewidth, clip, trim={0, 0.2cm, 0, 0.2cm}]{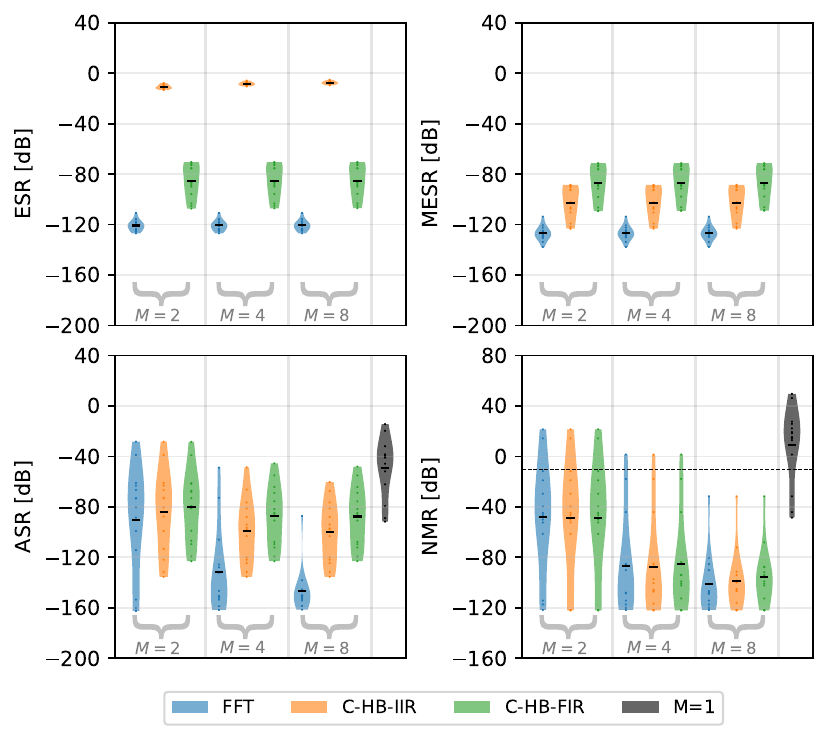}
    \caption{Oversampling experiment results -- output signal metrics for the 14 models with filenames containing \texttt{HighGain} in the GuitarML database. Each dot represents a different model with some dots overlapping. Lower is better in all cases. The ESR and MESR are averaged over input tone frequency; the ASR and NMR values are for $f_0=$ \SI{4186}{\Hz}. In the NMR plot, the dashed line indicates the \SI{-10}{\dB} threshold below which noise is assumed to be inaudible.}
    \label{fig:oversampling_violin}
\end{figure}

Here we expand the analysis to the remaining 14 models with filenames containing \texttt{HighGain} in the GuitarML database. Oversampling ratios of 2, 4, and 8 are considered and the performance of the FFT-based, C-HB-IIR, and C-HB-FIR methods is compared. Fig.~\ref{fig:oversampling_violin} shows the output signal metrics -- the ESR and MESR plots show the average over input frequency, whereas the aliasing metrics (ASR and NMR) are shown for $f_0=$ \SI{4186}{\Hz}. As expected, the FFT-based method gives the highest fidelity preservation of desired harmonic components. The IIR and FIR methods also perform well, both displaying an average MESR less than \SI{-70}{\dB} for all models. In terms of aliasing, it can be seen that at the training sample rate ($M=1$) the NMR at $f_0 =$ \SI{4186}{\Hz} exceeds the \SI{-10}{\dB} approximate threshold of audibility in 11 out of 14 models. Oversampling reduces the NMR progressively with $M=8$ reducing it below \SI{-10}{\dB} for all models. The NMR results at $f_0 =$ \SI{4186}{\Hz} are very similar regardless of resampling method, indicating that either the FIR or IIR method is suitable. Unless linear phase is desired, the IIR method is recommended as it involves fewer operations per sample and zero latency.

\section{Conclusions}\label{sec:conclusion}
This work investigated resampling methods for multirate neural audio processing, specifically focusing on two aspects: alleviating fractional sample rate discrepancies between the neural network training rate and the inference sample rate; and interpolation/decimation filter design for aliasing reduction via oversampling of RNNs.

For fractional discrepancies between the model rate and input signal rate, we examined the use of signal resampling before and after RNN processing, and found that this method could produce similar or better results to the previously proposed model adjustment methods (also referred to as sample rate independent RNNs or SRIRNNs), with many fewer filtering operations per sample. Specifically, it was found that Kaiser-window based FIR designs with \SI{120}{\dB} stopband attenuation, either as a single-stage resampling filter or cascaded with a half-band IIR interpolator/decimator, preserved the desired harmonic distortion components to a high fidelity without creating additional aliasing. Cascading with a half-band interpolator/decimator reduced the required transition bandwidth of the FIR filter and therefore latency to only \SI{0.1}{\ms} at the cost of phase distortion. 

We therefore conclude that the resampling method is likely preferable for most applications in audio effect processing unless zero phase delay and zero latency are required---in which case the model adjustment method may be a useful alternative. However, care must be taken when using the model adjustment method for under-sampling---when the inference signal rate is lower than the model training rate---as this is when instability is more likely \cite{Carson2025}. The resampling method avoids instability as the original RNN architecture is unmodified. 

For integer oversampling, interpolation and decimation filters were investigated to be used in conjunction with the model adjustment method. It was found that the proposed cascaded half-band interpolator methods -- both IIR and FIR -- outperform a previously proposed set of filters for oversampling by a factor of $M=8$. The half-band methods provided similar output quality to FFT-based resampling for $M=2, 4$, and $8$ times oversampling and can reduce aliasing in RNN models to inaudible levels, as measured by the noise-to-mask ratio. The C-HB-IIR method can be implemented with fewer operations per sample and zero latency and is therefore recommended over the C-HB-FIR method unless linear phase is desired. 

A perceptual evaluation of the proposed and baseline methods was not included, but this is an important area for future work. Specifically, the audibility of aliasing caused by RNN models of distortion effects should be investigated. A formal study into this may inform and encourage future research into anti-aliasing methods other than oversampling.

\section*{Acknowledgment}
This work began with Vesa Välimäki's research visit to the University of Edinburgh in October 2024. Thank you to the anonymous reviewers for their feedback on this article.
\ifCLASSOPTIONcaptionsoff
  \newpage
\fi


\bibliographystyle{IEEEtran}
\bibliography{refs}

\begin{IEEEbiography}[{\includegraphics[width=1in,height=1.25in,clip,keepaspectratio]{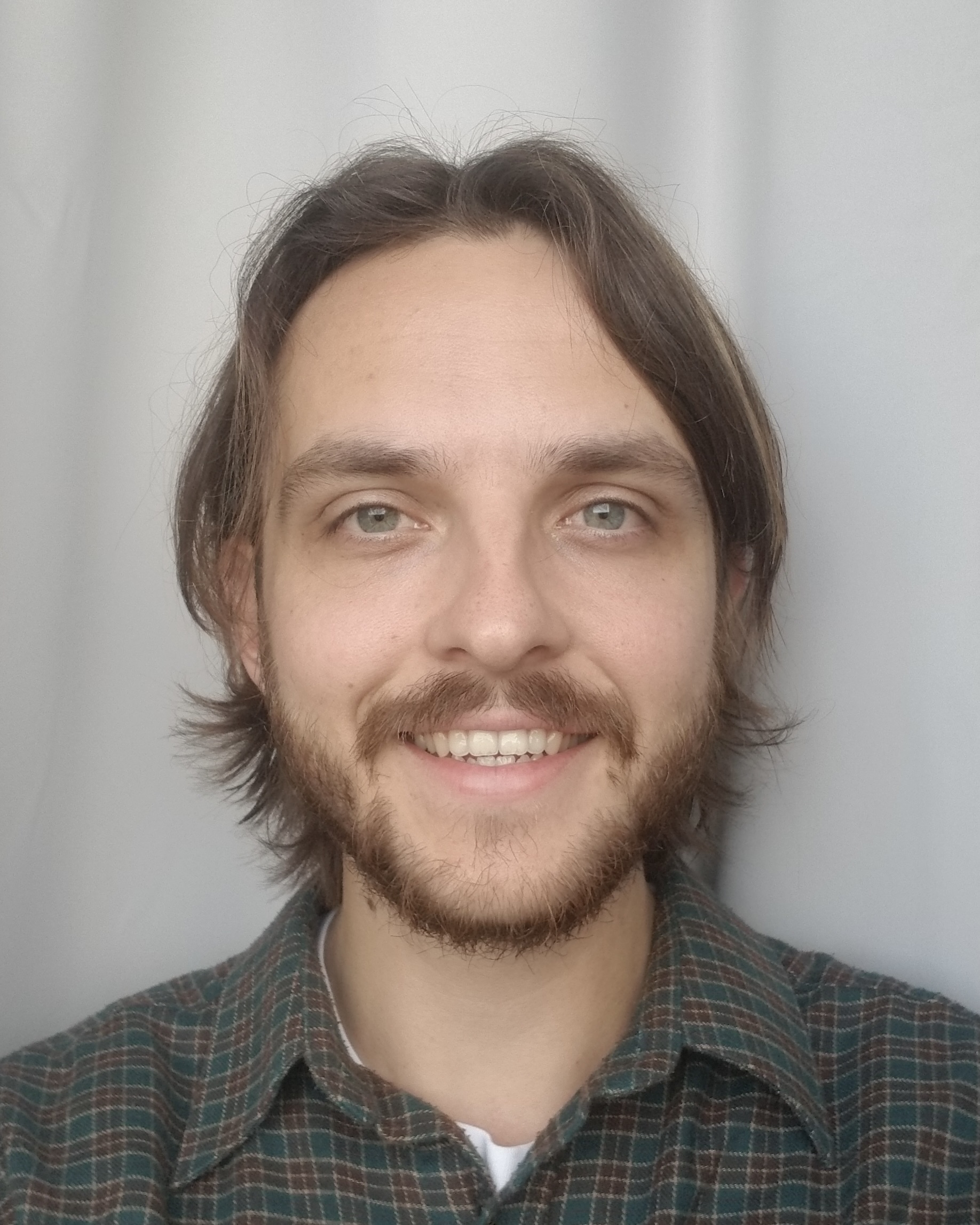}}]{Alistair Carson}
received his BEng (Hons) in Mechanical Engineering from the University of Strathclyde in 2018 and his MSc in Acoustics and Music Technology from the University of Edinburgh in 2020. Since then, he has been a member of the teaching staff in the Acoustics and Audio Group including a year-long post as Teaching Fellow in Acoustics. Currently, he is a doctoral researcher with interests in virtual analog modelling, differentiable DSP and deep learning for audio effect emulation.
\end{IEEEbiography}
\begin{IEEEbiography}
[{\includegraphics[width=1in,height=1.25in,clip,keepaspectratio]{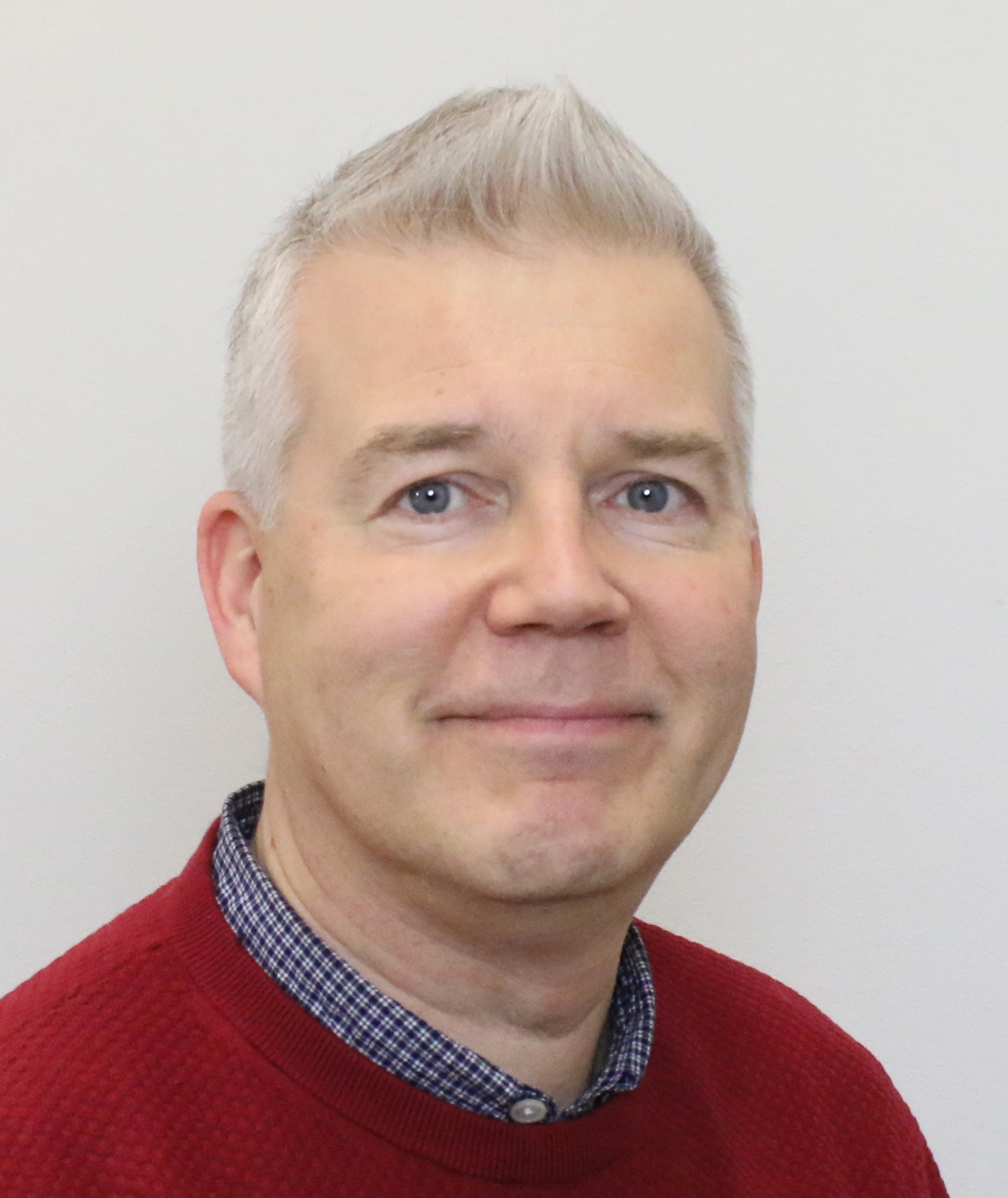}}]{Vesa V\"alim\"aki}
(Fellow, IEEE) received his MSc~and DSc~degrees in electrical engineering from the Helsinki University of Technology (TKK), Espoo, Finland, in 1992 and 1995, respectively.

He was a Postdoctoral Research Fellow at the University of Westminster, London, UK, in 1996. In 1997--2001, he was a Senior Assistant (cf.~Assistant Professor) at TKK. In 2001--2002, he was a Professor of signal processing at the Pori unit of the Tampere University of Technology. In 2008--2009, he was a Visiting Scholar at the Stanford University Center for Computer Research in Music and Acoustics. He is currently a Full Professor of audio signal processing and the Vice Dean for Research in electrical engineering at Aalto University, Espoo, Finland. His research interests are in audio and musical applications of signal processing and machine learning. 

Prof.~V\"alim\"aki is a Fellow of the IEEE, a Fellow of the Audio Engineering Society, and a Fellow of the Asia-Pacific Artificial Intelligence Association. In 2007--2013, he was a Member of the Audio and Acoustic Signal Processing Technical Committee of the IEEE Signal Processing Society. In 2005--2009, he served as an Associate Editor of the {\scshape IEEE Signal Processing Letters} and in 2007--2011, as an Associate Editor of the {\scshape IEEE Transactions on Audio, Speech and Language Processing}. In 2015--2020, he was a Senior Area Editor of the  {\scshape IEEE/ACM Transactions on Audio, Speech and Language Processing}. In 2007, 2015, and 2019, he was a Guest Editor of special issues of the {\scshape IEEE Signal Processing Magazine}, and in 2010, of a special issue of the {\scshape IEEE Transactions on Audio, Speech and Language Processing}. Since 2020, he has been the Editor-in-Chief of the {\it Journal of the Audio Engineering Society}.
\end{IEEEbiography}
\begin{IEEEbiography}
[{\includegraphics[width=1in,height=1.25in,clip,keepaspectratio]{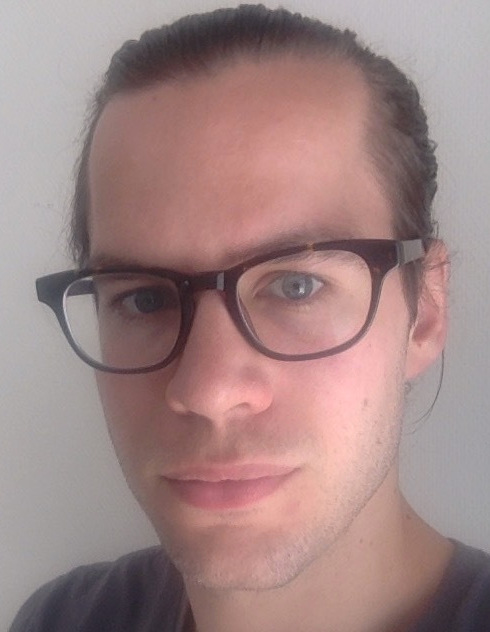}}]{Alec Wright}
is a Chancellor’s Fellow in Audio Machine
Learning in the Acoustics and Audio Group at the University of Edinburgh. He received his
D.Sc degree from Aalto University, Finland, in 2023 and his MSc in Acoustics and Music Technology from the University of Edinburgh in 2018. His
research interests include virtual analog modelling of guitar
amplifiers and other audio effects, and machine learning
for audio applications.
\end{IEEEbiography}
\begin{IEEEbiography}[{\includegraphics[width=1in,height=1.25in,clip,keepaspectratio]{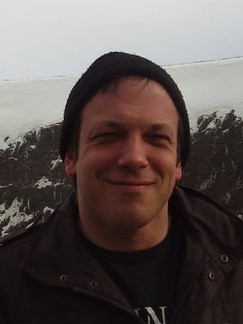}}]{Stefan Bilbao}
(B.A. Physics, Harvard, 1992, MSc., PhD Electrical Engineering, Stanford, 1996 and 2001 respectively) is currently Professor of Acoustics and Audio Signal Processing in the Acoustics and Audio Group at the University of Edinburgh, and previously held positions at the Sonic Arts Research Centre, at the Queen's University Belfast, and the Stanford Space Telecommunications and Radioscience Laboratory. He is an Associate Editor of JASA Express Letters, and was previously an associate editor of the IEEE/ACM Transactions on Audio, Speech and Language Processing. He works primarily on problems in acoustic simulation and audio signal processing. He was born in Montreal, Quebec, Canada. 
\end{IEEEbiography}

\end{document}